\pdfoutput=1
\documentclass[10pt,twocolumn,letterpaper]{article}

\usepackage[pagenumbers]{cvpr} 

\usepackage{graphicx}
\usepackage{amsmath}
\usepackage{amssymb}
\usepackage{booktabs}

\usepackage{algorithm}
\usepackage{algorithmic}
\usepackage{theorem}
            
\newtheorem{theorem}{Theorem} 
\newtheorem{definition}{Definition}

\newtheorem{lemma}{Lemma}

\usepackage{multirow}
\usepackage[T1]{fontenc} 

\usepackage[pagebackref,breaklinks,colorlinks]{hyperref}

\usepackage[capitalize]{cleveref}
\crefname{section}{Sec.}{Secs.}
\Crefname{section}{Section}{Sections}
\Crefname{table}{Table}{Tables}
\crefname{table}{Tab.}{Tabs.}

\def\cvprPaperID{3788} 
\def\confName{CVPR}
\def\confYear{2023}

\begin{document}

\title{SA-DPSGD: Differentially Private Stochastic Gradient Descent based on Simulated Annealing\\}

\author{Jie Fu, Zhili Chen, XinPeng Ling\\
East China Normal University, Shanghai, China\\
}

\maketitle

\begin{abstract}
   Differential privacy (DP) provides a formal privacy guarantee that prevents adversaries with access to machine learning models from extracting information about individual training points. Differentially private stochastic gradient descent (DPSGD) is the most popular training method with differential privacy in image recognition. However, existing DPSGD schemes lead to significant performance degradation, which prevents the application of differential privacy. In this paper, we propose a simulated annealing-based differentially private stochastic gradient descent scheme (SA-DPSGD) which accepts a candidate update with a probability that depends both on the update quality and on the number of iterations. Through this random update screening, we make the differentially private gradient descent proceed in the right direction in each iteration, and result in a more accurate model finally. In our experiments, under the same hyperparameters, our scheme achieves test accuracies 98.35\%, 87.41\% and 60.92\% on datasets MNIST, FashionMNIST and CIFAR10, respectively, compared to the state-of-the-art result of 98.12\%, 86.33\% and 59.34\%. Under the freely adjusted hyperparameters, our scheme achieves even higher accuracies, 98.89\%, 88.50\% and 64.17\%. We believe that our method has a great contribution for closing the accuracy gap between private and non-private image classification.
\end{abstract}

\begin{table*}[h]
\caption{This table summarizes the test accuracy of previous schemes and our scheme, SA-DPSGD, trained on the Mnist, FashionMNIST and CIFAR10 datasets. In our scheme, the values before the slash refer to the experimental results under the same hyperparameters as the previous art, and the values after the slash are the experimental results under the freely adjusted hyperparameters. For all experiments, we report the average accuracy across 5 independent runs.}
\centering
\begin{tabular}{|c|c|c|ccccc|}
\hline
\multirow{3}{*}{Dataset} & \multirow{3}{*}{Model} & \multirow{3}{*}{$(\epsilon,\delta)$} & \multicolumn{5}{c|}{Accuracy(\%)}                                                                                                                \\ \cline{4-8} 
                         &                & &\multicolumn{1}{c|}{DPSGD\cite{MartnAbadi2016DeepLW}} & \multicolumn{1}{c|}{{\begin{tabular}[c]{@{}c@{}}DPSGD\\ (tanh)\cite{NicolasPapernot2020TemperedSA}\end{tabular} }} & \multicolumn{1}{c|}{{\begin{tabular}[c]{@{}c@{}}DPSGD\\ (AUTO-S)\cite{ZhiqiBu2022AutomaticCD}\end{tabular} }} & \multicolumn{1}{c|}{{\begin{tabular}[c]{@{}c@{}}SA-DPSGD\\{[ours]}\end{tabular} }}       & {\begin{tabular}[c]{@{}c@{}}nonDP\\ $(\epsilon=\infty)$\end{tabular} } \\ \hline
Mnist        & 4-layer CNN               & (3,1e-5)                        & \multicolumn{1}{c|}{96.73} & \multicolumn{1}{c|}{98.05}       & \multicolumn{1}{c|}{98.12}         & \multicolumn{1}{c|}{\textbf{98.35/98.89}} & 99.09 \\ \hline
{FashionMNIST}      & 4-layer CNN         & (3,1e-5)                          & \multicolumn{1}{c|}{84.42} & \multicolumn{1}{c|}{86.03}       & \multicolumn{1}{c|}{86.33}         & \multicolumn{1}{c|}{\textbf{87.41/88.50}} & 89.01  \\ \hline
CIFAR10             & 9-layer CNN        & (3,1e-5)                          & \multicolumn{1}{c|}{52.06} & \multicolumn{1}{c|}{59.04}       & \multicolumn{1}{c|}{59.34}         & \multicolumn{1}{c|}{\textbf{60.92/64.17}} & 79.02 \\ \hline
\end{tabular}
\label{tab:comp}
\end{table*}

\section{Introduction}
\label{sec:intro}

\subsection{Background}
In the past decade, deep learning techniques have achieved remarkable success in various machine learning/data mining tasks, like medical image recognition\cite{AlexanderLundervold2018AnOO,SKevinZhou2020ARO,AndresJAnayaIsaza2021AnOO}.
Although we do not publish the image data when training the model, hidden adversaries may steal training data information by eavesdropping and analyzing the model parameters. For example, the contents of training data can be revealed if the trained models are attacked \cite{zhu2019deep,MattFredrikson2015ModelIA,MiladNasr2018ComprehensivePA,ZhiboWang2018BeyondIC,LeTrieuPhong2017PrivacyPreservingDL}, or the member information of the training dataset can be inferred if the threat models are trained and used \cite{song2017machine,LucaMelis2022ExploitingUF}. 
This is of particular concern in the field of computer vision, where many applications, such as medical imaging, require the handling of sensitive and legally protected data. The EU's General Data Privacy Regulation (GDPR) and the California Consumer Privacy Act\cite{cummings2018role} also require that machine learning practitioners need to take responsibility for protecting private data. One of the methods to prevent privacy disclosure in machine learning is the differential privacy (DP) technique \cite{dwork2014algorithmic}. The noise properly added to machine learning models published by a differentially private framework prevents unintentional leakage of private training data \cite{MartnAbadi2016DeepLW,NicholasCarlini2019TheSS,VitalyFeldman2019DoesLR,LiyaoXiang2019DifferentiallyPrivateDL,AnttiKoskela2018LearningRA,RezaShokri2015PrivacypreservingDL}.


\subsection{Prior Art}
Martín Abadi et al. \cite{MartnAbadi2016DeepLW}  proposed the first algorithm for deep learning with differential privacy, which is named DPSGD. Subsequently, many works aimed at improving the performance of DPSGD from different aspects. On the side of adaptive DPSGD, the work of \cite{AnttiKoskela2018LearningRA} used an adaptive learning rate to improve the convergence speed and reduce the privacy cost. Jaewoo Lee et al.\cite{JaewooLee2018ConcentratedDP} adaptively assigned different privacy budgets
to each training round to mitigate the effect of noise
on the gradient. Further, Zhiying Xu et al.\cite{ZhiyingXu2019AnAA} employed the Root Mean Square Prop (RMSProp) gradient descent technique to
adaptively add noise to coordinates of the gradient. Zhiqi Bu et al.\cite{ZhiqiBu2022AutomaticCD} proposed the AUTO-S clipping, which uses a small clipping norm by default and adds a positive stability constant $\gamma$ in the clipping gradient direction, that eliminates the need to tune 
clipping norm $C$ for any DP optimizers. For model structure, Nicolas Papernot et al.\cite{NicolasPapernot2020TemperedSA} founded that using a family of bounded activation functions (tempered sigmoids) instead of the unbounded activation function ReLU in DPSGD can achieve good performance. Anda Cheng et al.\cite{AndaCheng2021DPNASNA} proposed the framework that uses neural architecture search to automatically design models for private deep learning. Consequently, many works\cite{AdityaGolatkar2022MixedDP,DaYu2021DoNL,YingxueZhou2020BypassingTA} have focused on reducing the dimensionality of the model during training to reduce the impact of noise on the overall model. Last but no least, Florian Tramet et al.\cite{FlorianTramr2020DifferentiallyPL} used Scattering Network to traverse the image in advance to extract features before training can achieve high accuracy, but that is a pre-training method rather than a training from scratch. 

The above differentially private schemes for deep learning suffer from a significant utility loss, and actually there is still an obvious accuracy gap between differentially private deep learning and non-private techniques.

\subsection{Proposed Approach}

\begin{figure}[htbp]
	\begin{center}
		\includegraphics[width=0.9\linewidth]{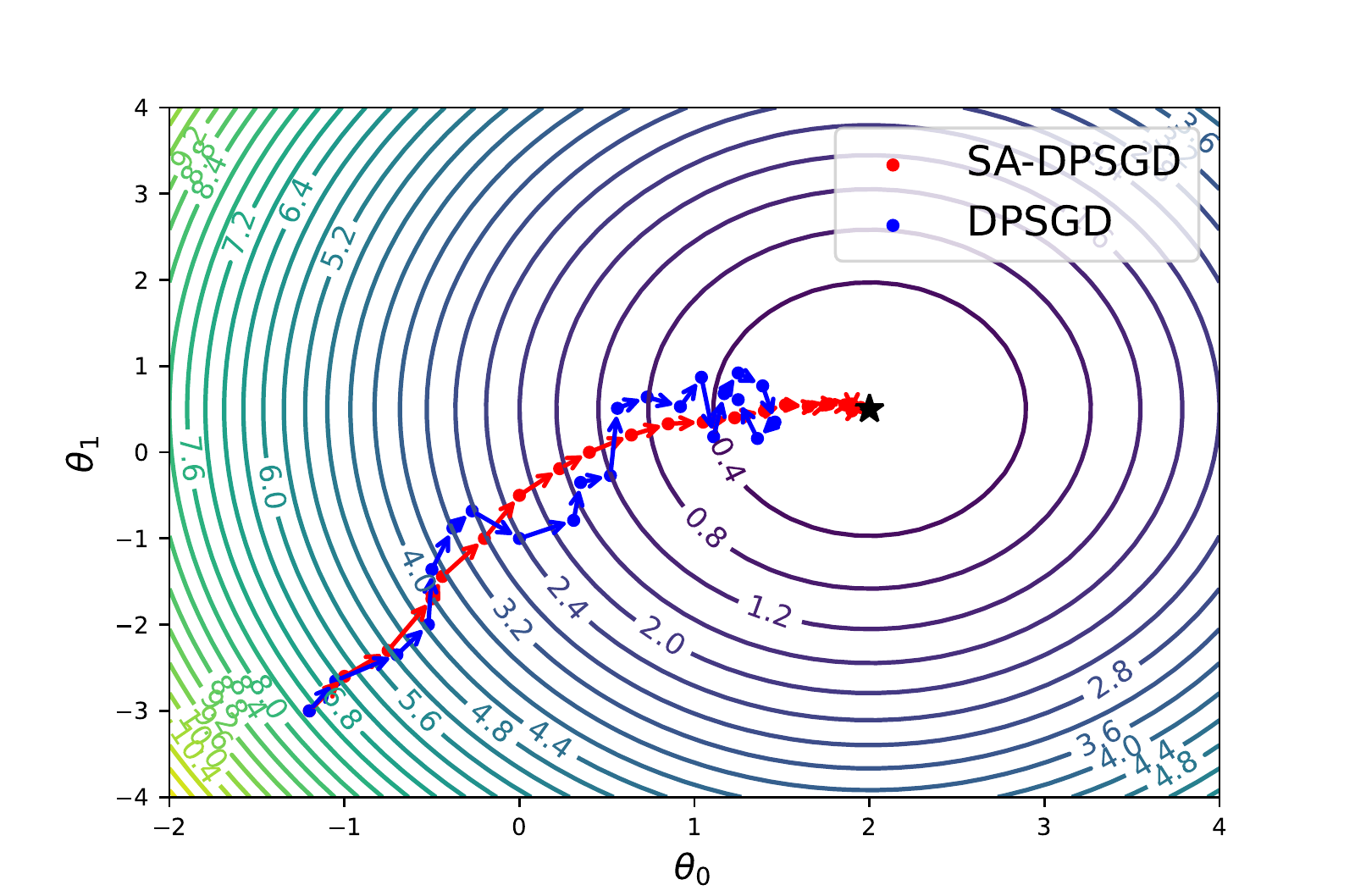}
		\caption{The simulated trajectories for DPSGD and SA-DPSGD schemes on linear regression.}
		\label{gra1}
	\end{center}
\end{figure}

Why do the existing schemes for differentially private deep learning still work not well enough? Our key observation is that in the stochastic gradient decent process these schemes allow all model updates no matter whether the corresponding objective function values get better. The fact is that in some model updates the objective function values may get worse after adding noise to gradients, especially when it is close to convergence (as shown by the blue trace line in Fig.~\ref{gra1}). This causes two consequences: worsening the optimization objective and wasting the privacy budget. Both consequences deteriorate the resulted models.
\vspace{-3mm}
\vspace{-5mm}

Based on the key observation, our main idea of improvement is to selectively perform model updates in the differentially private stochastic gradient decent. Specifically, we accept a model update if it produces a better objective function value, and reject it with a certain probability otherwise. We use probabilistic rejections other than deterministic ones, and, moreover, limit the number of continuous rejections, in order to avoid falling into a local optimum. This selection of model updates can make the differentially private gradient descent proceed in the right direction 
in each iteration, and result in a more accurate model finally (as shown by the red trace line in Fig.~\ref{gra1}).




Our main idea can be well implemented with the simulated annealing algorithm \cite{ScottKirkpatrick1983OptimizationBS}. The objective function value is the ``energy'', the number of iterations is the ``temperature'', a model update is a ``solution'', and we reject a solution with a certain probability concerning the energy change and the temperature. Experimental results listed in Table~\ref{tab:comp} show that our scheme outperforms significantly the state-of-the-art schemes DPSGD\cite{MartnAbadi2016DeepLW}, DPSGD(tanh)\cite{NicolasPapernot2020TemperedSA}, and DPSGD(AUTO-S)\cite{ZhiqiBu2022AutomaticCD}. When the hyperparameters are set as that used in the three schemes, our scheme achieves accuracies 98.35\%, 87.41\% and 60.92\% on image datasets MNIST, FashionMnist and CIFAR10, respectively, compared to the best result of the three schemes, 98.12\%, 86.33\%, and 59.34\%. When the hyperparameters are adjusted freely, our scheme achieves the best accuracies 98.89\%, 88.50\% and 64.17\%.

\subsection{Our Contribution}
In summary, our contributions are follows:
\begin{itemize}
    \item We propose an efficient scheme, SA-DPSGD, for differentially private deep learning, using the simulated annealing algorithm to select model updates with probability during the stochastic gradient decent process.
    \item We theoretically prove that SA-DPSGD satisfies differential privacy, and analyze its privacy loss with respect to the method of Rényi Differential Privacy (RDP).
    \item We fully implement the SA-DPSGD, and conduct extensive experiments on various datasets to demonstrate that it outperforms the state-of-the-art schemes significantly. 
\end{itemize}



\section{Preliminary Knowledge}
In this section, we briefly introduce the definition of differential privacy, the differentially private stochastic gradient descent (DPSGD) algorithm, and the simulated annealing algorithm.
\subsection{Differential Privacy}
Differential privacy is a rigorous mathematical framework that formally defines data privacy. Informally, it states that changes to a single data point in the input dataset cannot result in statistically significant changes in the output \cite{CynthiaDwork2006CalibratingNT,CynthiaDwork2011AFF,dwork2014algorithmic} if differential privacy holds.

\begin{definition}
	(Differential Privacy\cite{dwork2014algorithmic}). The randomized mechanism $A$ provides ($\epsilon$,  $\delta$)-Differential Privacy (DP), if for any two neighboring database $D$ and $D'$ that differ in only a single entry, $\forall$S $\subseteq$ Range($A$),
\end{definition}
\begin{equation}
{\rm Pr}(A(D) \in S) < e^{\epsilon} {\rm Pr}(A(D') \in S) + \delta
\end{equation}

Here, $\epsilon > 0$ gives the level of privacy guarantee in the worst case. The smaller $\epsilon$ is, the higher the privacy level is. The factor $\delta > 0$ allows for some probability that the property does not hold. In practice, this $\delta$ is required to be very small, and generally smaller than 1/$|D|$.

There are many attractive properties of DP. Sequential composition of DP will smoothly degrade privacy budget with multiple accesses to the same data. In addition, the output of a differentially private algorithm, can be arbitrarily processed without compromising the privacy guarantees, which is called the post-processing property.

\subsection{DPSGD}
Differentially Private Stochastic Gradient Descent (DPSGD), which was introduced by Abadi et al.\cite{MartnAbadi2016DeepLW}, is the first scheme for training deep neural networks with differential privacy guarantees. DPSGD adds noise to gradients in the training process, and results in models with differential privacy. Specifically, it clips each per-sample gradient according to a fixed $\ell_{2}$ norm, and adds Gaussian noise scaling with this norm to the sum of the gradients when computing the batch-averaged gradients. Then, the gradient descent is performed based on the batch-averaged gradients. Since initial models are randomly generated and independent on the sample data, and the batch-averaged gradients satisfy the differential privacy, the resulted models also satisfy the differential privacy due to the post-processing property. 


Originally, DPSGD employs moments accountant approach to calculate the privacy loss. In this work, for fair comparison, we all use the Rényi differential privacy (RDP) approach \cite{IlyaMironov2017RnyiDP,IlyaMironov2019RnyiDP} instead for privacy accountant. RDP is a generalization of differential privacy that uses the $\alpha$-Rényi divergences, and the RDP approach is implemented in Opacus \cite{AshkanYousefpour2021OpacusUD} and Tensorflow Privacy\cite{tensorflowPrivacy}. 


\subsection{Simulated Annealing Algorithm}

Simulated Annealing (SA) algorithm is a heuristic optimization algorithm proposed by Metropolis et al.\cite{NMetropolis1953EquationOS}. The idea of the algorithm was inspired by the physical annealing process, where when an object is cooled down, the probability of finding a low-energy particle increases gradually.



In the context of optimization, Kirkpatrick et al.\cite{ScottKirkpatrick1983OptimizationBS} regarded objective function value as the energy, and finding a better solution as finding a lower-energy particle. Assuming the objective function of the problem is $f(x)$, the current temperature is $T_i$, the current feasible solution is $x_i$, and the new solution is $x^{\prime}$, according to the simulated annealing process, the probability that the new solution $x^{\prime}$ is accepted as the next feasible solution $x_{i+1}$ is designed as Eq.~\eqref{equ:probability}, and the simulated annealing algorithm is depicted in Algorithm~\ref{alg:1}.
\begin{equation}\label{equ:probability}
    P\left(x^{\prime} \rightarrow x_{i+1}\right)=\left\{\begin{array}{cc}
1 , & f\left(x^{\prime}\right) \leq f\left(x_{i}\right) \\
e^{-\frac{f\left(x^{\prime}\right)-f\left(x_{i}\right)}{T_{i}}}, & f\left(x^{\prime}\right)>f\left(x_{i}\right)
\end{array}\right.
\end{equation}

\begin{algorithm}
	\renewcommand{\algorithmicrequire}{\textbf{Input:}}
	\renewcommand{\algorithmicensure}{\textbf{Output:}}
	\caption{Simulated Annealing Algorithm}
	\label{alg:1}
	\begin{algorithmic}[1]
	\REQUIRE {Objective function $f$ to be minimized, initial temperature $T_0>0$, cooling factor $\alpha \in (0,1)$, number of iterations $n$}
        \STATE Initialize: initial solution $s^{(0)}$, $T=T_0$
		\FOR{$i=0,...,n-1$}
		\STATE Pick a new solution $s_{\emph{new}}$ in a neighborhood of $s^{(i)}$
        \STATE $\triangle f=f(s_{\emph{new}})-f(s^{(i)})$
        \STATE $\emph{prob}=e^{-\triangle f/T}$
        \IF{$\operatorname{random}(0,1)<\emph{prob}$}
        \STATE $s^{(i+1)}=s_{\emph{new}}$
        \ELSE
        \STATE $s^{(i+1)}=s^{(i)}$
        \ENDIF
        \STATE $T=\alpha \cdot T$
        \ENDFOR
		\STATE \textbf{Output} the last solution $s^{(n-1)}$
	\end{algorithmic}  
\end{algorithm}




\section{SA-DPSGD}




In this section, we first give the overview of SA-DPSGD, then describe its detailed design, and finally provide the privacy analysis.

\subsection{Overview}

We combine our main idea of selecting model updates with the simulated annealing algorithm, and design the SA-DPSGD scheme. The high-level description of our scheme is as follows.
\begin{itemize}
\item We generate new solutions (updates) iteratively by DPSGD, and compute the energy (objective function value) accordingly. The acceptance probability of the current solution is then determined by the energy change from the previous iteration to the current one, and the number of accepted solutions by far. Note that we use the number of accepted solutions instead of that of all generated solutions to more accurately reflect the effectiveness of actual iterations.
\item We let the acceptance probability be always 1, when the energy change is negative. That is, we always accept the solutions (updates) that step in the right direction. Even though our model updates are noisy, which means that the actual energy may be positive with a tiny probability, this still guarantees that the training proceeds mainly in the direction to convergence.
\item We let the acceptance probability decreases exponentially as the energy change and the number of accepted solutions increase, when the energy change is positive. In this case, the energy would become worse if a solution were accepted. However, deterministic rejections may cause the final solution to fall in a local optimum. Therefore, we accept solutions of positive energy changes with a small, decreasing probability.
\item We limit the number of continuous rejections, and accept a solution anyway if too many successive rejections take place previously. When the training is close to convergence, the acceptance probability may become so small that it nearly rejects all solutions with positive energy changes, and may reach a local optimum. Limitation of the continuous rejections avoids this problem by accepting a solution anyway when necessary.
\end{itemize}

\begin{figure}[htbp]
	\begin{center}
		\includegraphics[width=1.0\linewidth]{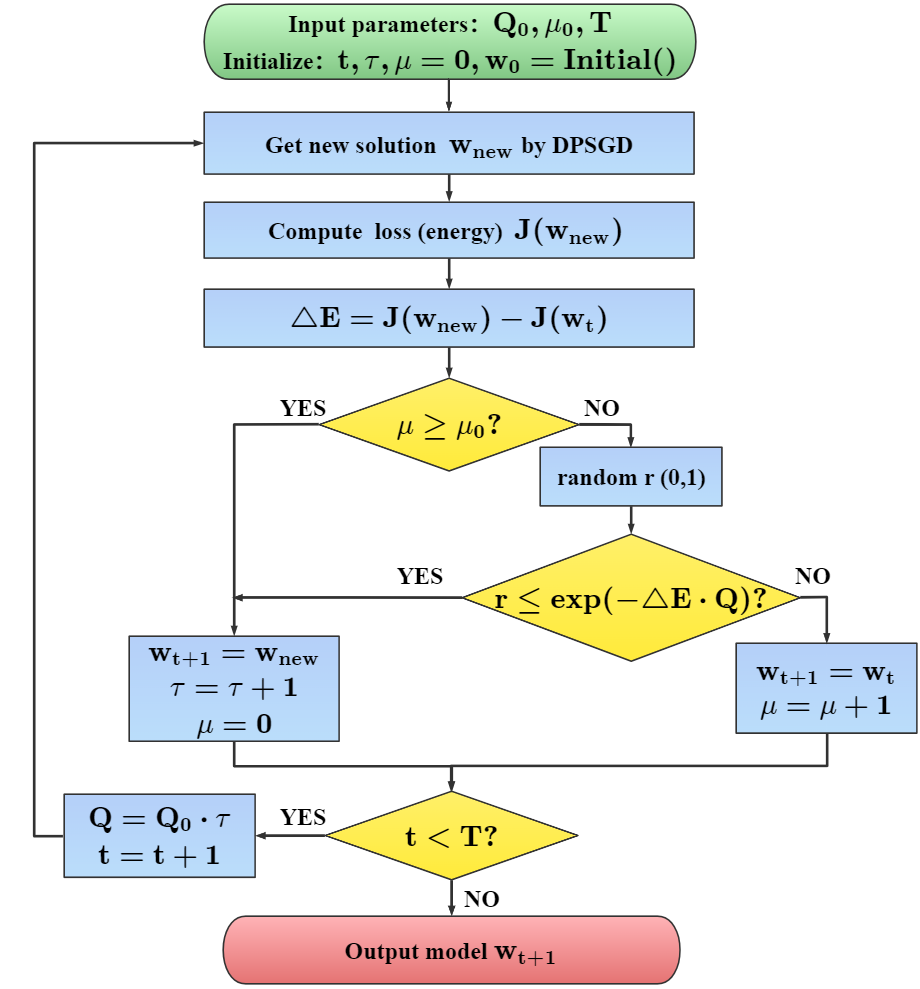}
		\caption{The work flow of SA-DPSGD.}
		\label{gra2}
	\end{center}
\end{figure}

Following the above high-level description, we depict the work flow of SA-DPSGD as shown in Fig.~\ref{gra2}. First, all the parameters are initialized. Given the initial temperature $Q_0$, the rejection threshold $\mu_0$, and the total number of iterations $T$, the number of iterations $t$, the number of accepted solutions $\tau$, and the number of rejections $\mu$ are all set to 0, and the initial model $w_0$ is randomly generated. Next, we use the cross-entropy loss function as the objective function, and in the current iteration (assuming the $(t+1)$-th iteration), a new solution $w_{new}$, its energy $J(w_{new})$, and the energy change $\Delta E$ from the previous iteration to the current one, are computed using DPSGD. Then, the neural network determines whether to accept the new solution. If the number of continuous rejections $\mu$ exceeds the threshold value $\mu_0$, it accepts the solution anyway, i.e., $w_{t+1} = w_{new}$; otherwise, it accepts the solution with a probability $P$ as Eq.~\eqref{equ:probability2}, where $Q$ is the current temperature.

\begin{equation}\label{equ:probability2}
P=\left\{\begin{array}{cc}
1, & \Delta E \leq 0 \\
e^{-\Delta E*Q}, & \Delta E > 0
\end{array}\right.    
\end{equation}







Note that, SA-DPSGD differs from conventional simulated annealing algorithm in the definition of temperature. We define the temperature $Q = Q_0 \cdot \tau$, where $Q_0$ is the initial temperature and $\tau$ is the number of accepted solutions by far. Therefore, our temperature increases instead as the training proceeds. However, we compute the acceptance probability by multiplying instead of dividing the temperature. This makes the probability decrease as the training goes, which is just the same as the conventional simulated annealing algorithm.

\vspace{-5mm}
\subsection{Detailed Design}
We now present the detailed design of SA-DPSGD in Algorithm~\ref{alg:2}, which consists of the following steps.


\begin{itemize}
	\item \textbf{Step0: Initialization (Line~\ref{line:init}).} Initialize parameters ($\mu_0$, $Q_0$, $w_0$, etc.)
	\item \textbf{Step1: Gradient Computation (Lines~\ref{line:sample} and \ref{line:gradient}).} Randomly select a small batch of samples $B_t$ from the training dataset, and for each sample $i \in B_t$, calculate the corresponding gradient values.
	\item \textbf{Step2: Gradient Clipping (Line~\ref{line:clip}).} Clip the gradients of each sample so that the $l_2$ norm of the gradients is not greater than the clipping norm $C$.
	\item \textbf{Step3: Noise Adding (Line~\ref{line:noise}).} First sum the clipped gradients, then add Gaussian noise $\mathcal{N}(0,C^2\sigma^2)$ to the sum, and finally compute the average.
	\item \textbf{Step4: Gradient Decent (Line~\ref{line:decent}).} Perform gradient descent using the noisy gradients to obtain a new solution (model update) $w_{new}$ for the current iteration.
	\item \textbf{Step5: Evaluating New Solution (Lines~\ref{line:loss} to \ref{line:endif}).} Compute the energy (i.e., objective function value) $J(w_{new})$ of the new solution, and subtract the energy of the previous iteration from it to get the energy change $\Delta E$. The new solution is accepted if the number of rejections exceeds the rejection threshold, i.e., $\mu \ge \mu_0$; otherwise, it is accepted with a probability $P$ base on Eq.~\eqref{equ:probability2}.
	\item \textbf{Step6: Raising the temperature (Lines~\ref{line:temperature}).} We let the current temperature $Q=Q_0 \cdot \tau$, so that the temperature rises periodically to produce a lower acceptance probability as the training proceeds.
	\item  Repeat steps 1-6 until the threshold $T$ for the total number of training sessions is reached. Finally, the model $w_{t+1}$ is output and the privacy loss is calculated according to Theorem 1.
\end{itemize}

\begin{algorithm}
	\renewcommand{\algorithmicrequire}{\textbf{Input:}}
	\renewcommand{\algorithmicensure}{\textbf{Output:}}
	\caption{SA-DPSGD}
	\label{alg:2}
	\begin{algorithmic}[1]
	\REQUIRE {Samples $\left\{x_{1}, \ldots, x_{N}\right\}$, loss function $\mathcal{L}(\theta, x)$. Parameters: total number of iterations $T$, learning rate $\eta$, noise scale $\sigma$, batch size $B$, clipping norm $C$, rejection threshold $\mu_0$, initial temperature $Q_0$}
		\STATE Initialize: $t,\tau,\mu=0$, $Q=Q_0$, $w_0$ \label{line:init}
		\WHILE {$t<T$}
		\STATE Sample randomly a batch $B_t$ with probability B/N \label{line:sample}
        \FOR{$i\in B_t$}
		\STATE Compute $g_t(x_i)\leftarrow \nabla \mathcal{L}{(w_t,x_i)}  $ \label{line:gradient}
		\STATE $\overline{g}_t(x_i) \leftarrow g_t{(x_i)} / max(1,\frac{||g_t{(x_i)}||_2}{C}) $ \label{line:clip}
        \ENDFOR
		\STATE $\widetilde{g}_t \leftarrow \frac{1}{B} (\sum_{i}^{B_t} \overline{g}_t(x_i)+\mathcal{N}(0,\sigma^2 {C}^2))$\label{line:noise}
		\STATE $w_{new}=w_t - \eta_t \widetilde{g}_t$ \label{line:decent}
		\STATE Compute test loss $J(w_{new})$
		\STATE $\triangle E=J(w_{new})-J(w_{t})$\label{line:loss}
		\STATE $P=e^{-\triangle E \cdot Q}$
		\IF {$ random(0,1) \le P \enspace or \enspace \mu \ge \mu_0 $}
		\STATE $w_{t+1} = w_{new}$
		\STATE $\tau=\tau+1$	
		\STATE $\mu=0$
		\ELSE
		\STATE $w_{t+1} = w_{t}$
		\STATE $\mu=\mu+1$
		\ENDIF\label{line:endif}
		\STATE $t=t+1$
		\STATE $Q=Q_0 \cdot \tau$ \label{line:temperature}
    \ENDWHILE
		\STATE \textbf Compute privacy loss $(\epsilon,\delta)$ by Theorem 1
		\STATE \textbf{Output} $w_{t+1}$ and privacy loss $(\epsilon,\delta)$
	\end{algorithmic}  
\end{algorithm}

\subsection{Privacy Analysis}
In SA-DPSGD as shown in Algorithm~\ref{alg:2}, initial models are randomly generated not depending on sample data, the batch-averaged gradients are added noise to satisfy differential privacy, and all other operations including model selecting and model updating are only based on the initial models and the batch-averaged gradients. Thus, SA-DPSGD satisfies differential privacy due to the post-processing property.

We claim that the privacy loss caused by SA-DPSGD is only dependent on the number of accepted model updates $\tau$, since when final models are published, only the results of accepted model updates can be seen, and whether resampling batches and re-generating noise are simply unknown. Theorem~\ref{the:privacy-loss} shows the privacy loss calculation of SA-DPSGD.


\begin{theorem}(DP Privacy Loss of SA-DPSGD). After $\tau$ iterations of model gradient decent, the privacy loss of SA-DPSGD satisfies: \label{the:privacy-loss}
\begin{equation}
\begin{split}
    (\epsilon,\delta)=( \frac{\tau}{\alpha-1}\sum_{i=0}^{\alpha}\left(\begin{array}{l}
\alpha \\ i
\end{array}\right)(1-q)^{\alpha-i} q^{i}\\ \exp \left(\frac{i^{2}-i}{2 \sigma^{2}}\right)
+ \frac{\log 1/\delta}{\alpha-1},\delta)
\end{split}
\end{equation}
where $q=\frac{B}{N}$ , $\sigma$ is noise scale, and $\alpha > 1$ is the order.
\end{theorem}
$Proof.$ We calculate the privacy loss based on RDP approach\cite{IlyaMironov2017RnyiDP}. We first use the sampling Gaussian theorem of RDP to calculate the privacy cost of each iteration, then use the composition of RDP mechanisms to compute the privacy cost of multiple iterations, and finally convert the obtained RDP cost to DP cost.

Definitions~\ref{def:sgm} and \ref{def:rdp} define Sampled Gaussian Mechanism (SGM) and Rényi Differential Privacy (RDP), respectively.
\begin{definition}\label{def:sgm}
(Sampled Gaussian Mechanism (SGM)\cite{IlyaMironov2019RnyiDP}). Let $f$ be a function mapping subsets of $S$ to $\mathbb{R}^d$. We define the Sampled Gaussian Mechanism (SGM) parameterized with the sampling rate $0 < q \leq 1$ and the  $\sigma > 0$ as
\begin{equation}
\begin{aligned}
	S G_{q, \sigma}(S) \triangleq & f(\{x: x \in S \text { is sampled with probability } q\}) \\
	&+\mathcal{N}\left(0, \sigma^{2} \mathbb{I}^{d}\right)
	\end{aligned}
\end{equation}
in SA-DPSGD, $f$ is the clipped gradient evaluation on sampled data points $f(\{x_i\}_{i\in B}) = \sum_{i\in B} \overline{g}_t(x_i)$. If $ \overline{g}_t$ is obtained by clipping $g_t$ with a gradient norm bound $C$, then the sensitivity of $f$ is equal to $C$.
\end{definition}

\begin{definition}\label{def:rdp}
(RDP privacy budget of SGM\cite{IlyaMironov2019RnyiDP}). Let $SG_{q,\sigma}$, be the Sampled Gaussian Mechanism for some function $f$. If $f$ has sensitivity 1, $SG_{q,\sigma}$ satisfies $(\alpha,\epsilon)$-RDP whenever
\begin{equation}
\epsilon \leq \frac{1}{\alpha-1} \log max(A_{\alpha}(q,\sigma),B_{\alpha}(q,\sigma))
\end{equation}
where
\begin{equation}
\left\{\begin{array}{l}
A_{\alpha}(q, \sigma) \triangleq \mathbb{E}_{z \sim \mu_{0}}\left[\left(\mu(z) / \mu_{0}(z)\right)^{\alpha}\right] \\
B_{\alpha}(q, \sigma) \triangleq \mathbb{E}_{z \sim \mu}\left[\left(\mu_{0}(z) / \mu(z)\right)^{\alpha}\right]
\end{array}\right.
\end{equation}
with $\mu_{0} \triangleq \mathcal{N}\left(0, \sigma^{2}\right), \mu_{1} \triangleq \mathcal{N}\left(1, \sigma^{2}\right) \mbox { and } \mu \triangleq(1-q) \mu_{0}+q \mu_{1}$.

Furthermore, it holds $\forall(q,\sigma)\in(0,1] \times \mathbb{R}^{+},A_{\alpha}(q,\sigma) \geq B_{\alpha}(q, \sigma) $. Thus, $ S G_{q, \sigma}$  satisfies  $\left(\alpha, \frac{1}{\alpha-1} \log \left(A_{\alpha}(q, \sigma)\right)\right)$-RDP .

Finally, paper \cite{IlyaMironov2019RnyiDP} describes a procedure to compute $A_{\alpha}(q,\sigma)$ depending on integer $\alpha$ as Eq.~\eqref{equ:a-alpha}.
\begin{equation}\label{equ:a-alpha}
A_{\alpha}=\sum_{k=0}^{\alpha}\left(\begin{array}{l}
\alpha \\ k
\end{array}\right)(1-q)^{\alpha-k} q^{k} \exp \left(\frac{k^{2}-k}{2 \sigma^{2}}\right)
\end{equation}
\end{definition}

Lemma~\ref{lem:composition} gives the composition of RDP mechanisms.
\begin{lemma}\label{lem:composition}
(Composition of RDP\cite{IlyaMironov2017RnyiDP}). For two randomized mechanisms $f, g$ such that $f$ is $(\alpha,\epsilon_1)$-RDP and $g$ is $(\alpha,\epsilon_2)$-RDP the composition of $f$ and $g$ which is defined as $(X, Y )$(a sequence of results), where $ X \sim f $ and $Y \sim g$, satisfies $(\alpha,\epsilon_1+\epsilon_2)-RDP$
\end{lemma}\label{lem:iterations}

According to Definitions~\ref{def:sgm} and \ref{def:rdp}, and Lemma~\ref{lem:composition}, we have Theorem~\ref{the:compositon}.
\begin{theorem}\label{the:compositon} Given the sampling rate $q=\frac{B}{N}$ for each iteration of the local dataset and the noise factor $\sigma$ for all iterations, the total RDP privacy loss for $T$ accepted iterations for any integer $\alpha \geq 2$ is given in Eq~\eqref{equ:epsilon}.
\begin{equation}\label{equ:epsilon}
    \epsilon^{'}(\alpha)_T= \frac{T}{\alpha-1}\sum_{i=0}^{\alpha}\left(\begin{array}{l}
\alpha \\ i
\end{array}\right)(1-q)^{\alpha-i} q^{i} \exp \left(\frac{i^{2}-i}{2 \sigma^{2}}\right)
\end{equation}
\end{theorem}

Lemmas~\ref{lem:conversion} and \ref{lem:bconversion} give two ways of conversion from RDP to DP, and the latter is more compact. However, since the previous work used Lemma~\ref{lem:conversion} for calculating
privacy loss, for fair comparison we also uses it in this work. 

\begin{lemma}\label{lem:conversion}
(Conversion from RDP to DP\cite{IlyaMironov2017RnyiDP}). if a randomized mechanism $f : D \rightarrow \mathbb{R}$  satisfies $(\alpha,\epsilon)$-RDP ,then it satisfies$(\epsilon +\frac{\log 1/\delta}{\alpha-1},\delta)$-DP where $0<\delta<1$
\end{lemma}

\begin{lemma}\label{lem:bconversion}
(Better conversion from RDP to DP\cite{balle2020hypothesis}). If a mechanism M is $(\alpha,\epsilon)$-RDP, then it is $ (\epsilon+\log ((\alpha-1) / \alpha)-(\log \delta+ \log \alpha) /(\alpha-1), \delta)$ for any $0<\delta<1$.
\end{lemma}

By Theorem~\ref{the:compositon} and Lemma~\ref{lem:conversion}, Theorem~\ref{the:privacy-loss} is proved. $\Box$

In practice, given $\sigma$, $\delta$ and $B$ at each iteration, we select $\alpha$ from $\left\{2,3,...,64\right\}$ and determine the smallest  $\epsilon_{*}$ in Theorem~\ref{the:privacy-loss}. The privacy loss is the pair $(\epsilon_{*},\delta)$.





\section{Experiment}
We evaluate SA-DPSGD on three datasets: MNIST, FashionMNIST and CIFAR10. Although the three datasets are regarded as ``solved'' in the computer vision community\cite{KaimingHe2015DeepRL,GaoHuang2016DenselyCC}, they still remain challenging in the context of differential privacy \cite{EugeneBagdasaryan2019DifferentialPH,papernot2019machine,AshkanYousefpour2021OpacusUD}.


\subsection{Initialization}
\textbf{Dataset.} 
MNIST\cite{LiDeng2012TheMD} is the standard dataset for handwritten digit recognition and FashionMNIST\cite{HanXiao2017FashionMNISTAN} is the dataset for clothing classification, both of them consist of 60,000 training examples and 10,000 testing examples. Each example is a 28 $\times$ 28 gray-level image.
CIFAR10\cite{AlexKrizhevsky2009LearningML} consists of color images classified into 10 classes, and partitioned into 50,000 training examples and 10,000 test examples. Each example is a 32 $\times$ 32 image with three channels (RGB).

\textbf{Network model.} 
For MNIST and FashionMNIST, we use a simple convolutional neural network with 4 layers, and for CIFAR10, we use a deeper convolutional neural network model with 9 layers. The network architectures are all the same as previous work \cite{FlorianTramr2020DifferentiallyPL,ZhiqiBu2022AutomaticCD,NicolasPapernot2020TemperedSA}.
 


\textbf{Parameter settings.}
Our experiments are implemented with SGD optimizer by pytorch\cite{pytorch2018pytorch}. All hyperparameter settings for this experiment, detailed in Table~\ref{table:2}, are the same as \cite{FlorianTramr2020DifferentiallyPL,ZhiqiBu2022AutomaticCD}, except the initial temperature $Q_0$ and rejection threshold $\mu_0$ are new parameters in our scheme.
\begin{table}[htb]
\centering
\caption{The setting of hyperparameters}
\begin{tabular}{lccc}
\toprule
Parameter & \text {MNIST} & \begin{tabular}[c]{@{}c@{}}Fashion\\ MNIST\end{tabular} & \begin{tabular}[c]{@{}c@{}}CIFA\\ R10\end{tabular}\\
\midrule
Learning rate $\eta$ & 0.5 & 4.0 & 1.0\\
Batch size $B$ & 512 & 2048 & 1024\\
Clipping norm $C$ & 0.1 & 0.1 & 0.1\\
Noise scale $\sigma$ & 1.23 & 2.15 & 1.54\\
Privacy budget $(\epsilon,\delta)$ & (3,1e-5) & (3,1e-5) & (3,1e-5)\\
Initial temperature $Q_0$ & 10.0 & 10.0 & 10.0\\
Rejection threshold $\mu_0$ & 10 & 10 & 5\\
\bottomrule
\end{tabular}

\label{table:2}
\end{table}

\textbf{Baselines and prior state-of-the-art.}
We evaluate the privacy cost and accuracy of our approach SA-DPSGD compared with state-of-the-art schemes: DPSGD\cite{MartnAbadi2016DeepLW}, DPSGD(tanh)\cite{NicolasPapernot2020TemperedSA} and DPSGD(AUTO-S)\cite{ZhiqiBu2022AutomaticCD}. They are all differentially private deep learning schemes from scratch, where DPSGD uses ReLU as the activation function of the neural network by default, DPSGD(tanh) and DPSGD(AUTO-S) apply tanh activation function instead. To the best of our knowledge, DPSGD(AUTO-S) performs best among the existing differentially private deep learning methods from scratch. Our scheme selects tanh as the activation function, and all schemes apply the above network structure and parameter settings.

\subsection{Privacy Cost}


We compare the privacy cost of our scheme, SA-DPSGD, with that of the state-of-the-art schemes, when achieving the same accuracy on datasets MNIST, FashionMNIST and CIFAR10. The experimental results are summarized in Table~\ref{table:3}, where $\epsilon_{D}$, $\epsilon_{H}$, $\epsilon_{A}$ and $\epsilon_{S}$ denote the minimum privacy cost of  DPSGD, DPSGD(tanh), DPSGD(AUTO-S) and SA-DPSGD, respectively. The parameter $\delta$ is set to $10^{-5}$, and RDP approach is used for privacy loss calculation for all the schemes. We observe that SA-DPSGD always has a lower privacy cost than other schemes, and the higher the privacy costs are, the bigger the gap of privacy costs is. The reason should be that our scheme accepts right model updates, and rejects wrong ones, which speeds up the convergence while saves the privacy cost, and when the privacy costs are higher, the saving is more obvious. 

\begin{table}[h]
\centering
\caption{This table summarizes the privacy cost of previous works and our scheme SA-DPSGD trained on datasets MNIST, FashionMNIST and CIFAR10 when a pre-specified level of test accuracy was
achieved. For all experiments, we report the average privacy loss across 5 independent runs. "-" indicates that the corresponding accuracy cannot be achieved within $\epsilon \leq 6$.}
\begin{tabular}{lccccc}		
\toprule
Dataset & Accuracy &  $\epsilon_D$ & $\epsilon_H$ & $\epsilon_A$ & $\epsilon_S$\\
\midrule
\multirow{3}*{MNIST} & 0.94  & 1.64 & 1.23 & 1.22 & \textbf{1.21} \\
~ & 0.96  & 2.47 & 1.47 & 1.47 & \textbf{1.46}\\
~ & 0.98  & - & 2.92 & 2.91 & \textbf{2.53}\\
\midrule
\multirow{3}*{\begin{tabular}[c]{@{}c@{}}Fashion\\ MNIST\end{tabular} }& 0.82   & 2.20 & 1.73  & 1.64 & \textbf{1.52}\\
~ & 0.84 & 2.88 & 2.08 & 2.07 & \textbf{1.86}\\
~ & 0.86 & 5.86 & 2.96 & 2.92 & \textbf{2.51}\\
\midrule
\multirow{3}*{CIFAR10} & 0.52 & 2.99 & 2.15  & 2.05 & \textbf{1.93}\\
~ & 0.55  & 3.68 & 2.39 & 2.36 & \textbf{2.27}\\
~ & 0.58  & 4.54 & 2.83 & 2.80 & \textbf{2.57}\\
\bottomrule
\end{tabular}

\label{table:3}
\end{table}

\vspace{-3mm}
\subsection{Performace}

\begin{figure*}[h]
	\centering
	\begin{subfigure}{0.32\linewidth}
		\centering
		\includegraphics[width=1.0\linewidth]{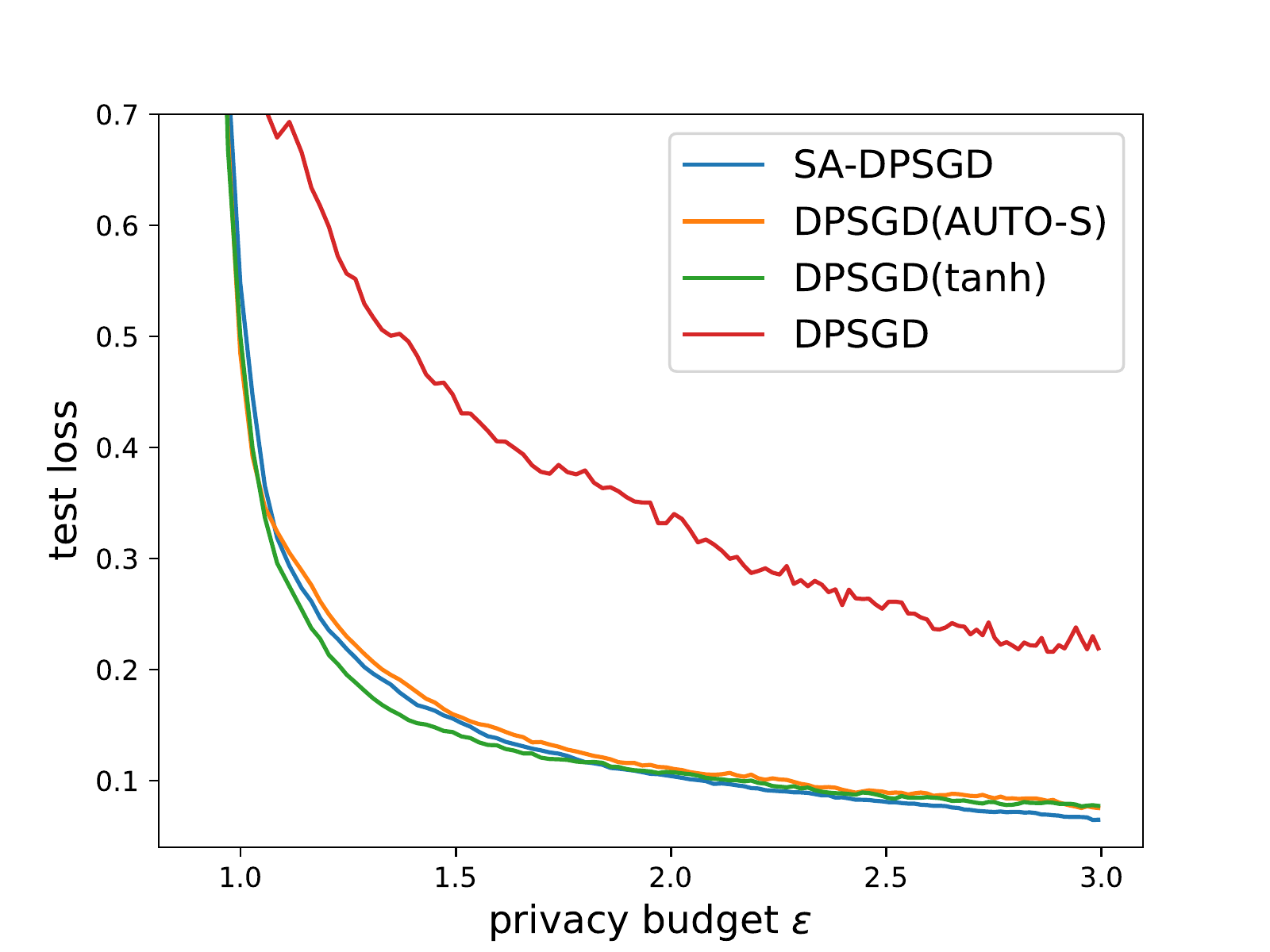}
		\caption{Avg test loss result in MNIST}
	\end{subfigure}
	\centering
	\begin{subfigure}{0.32\linewidth}
		\centering
		\includegraphics[width=1.0\linewidth]{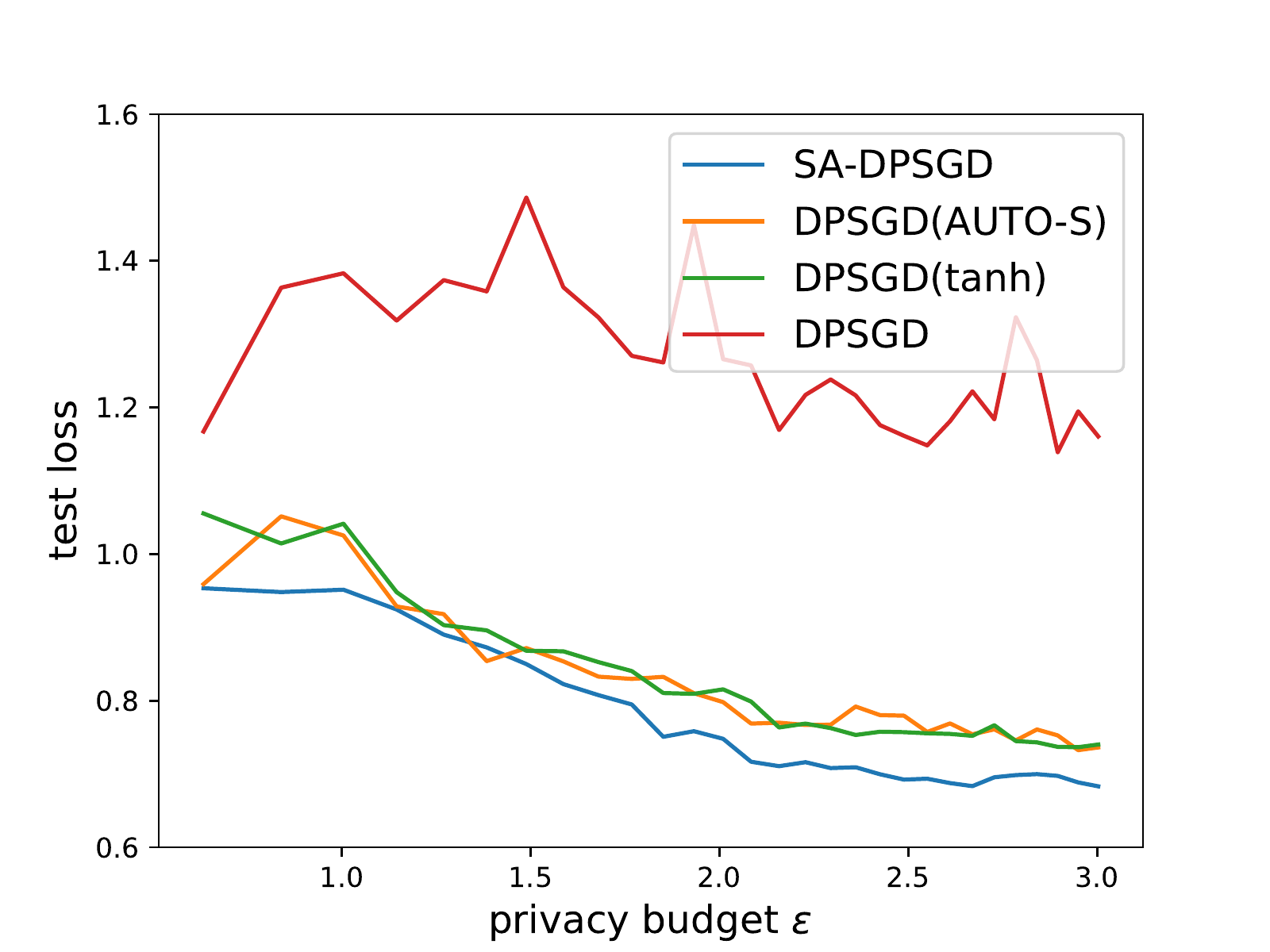}
		\caption{Avg test loss result in FashionMMNSIT}
	\end{subfigure}
	\centering
	\begin{subfigure}{0.32\linewidth}
		\centering
		\includegraphics[width=1.0\linewidth]{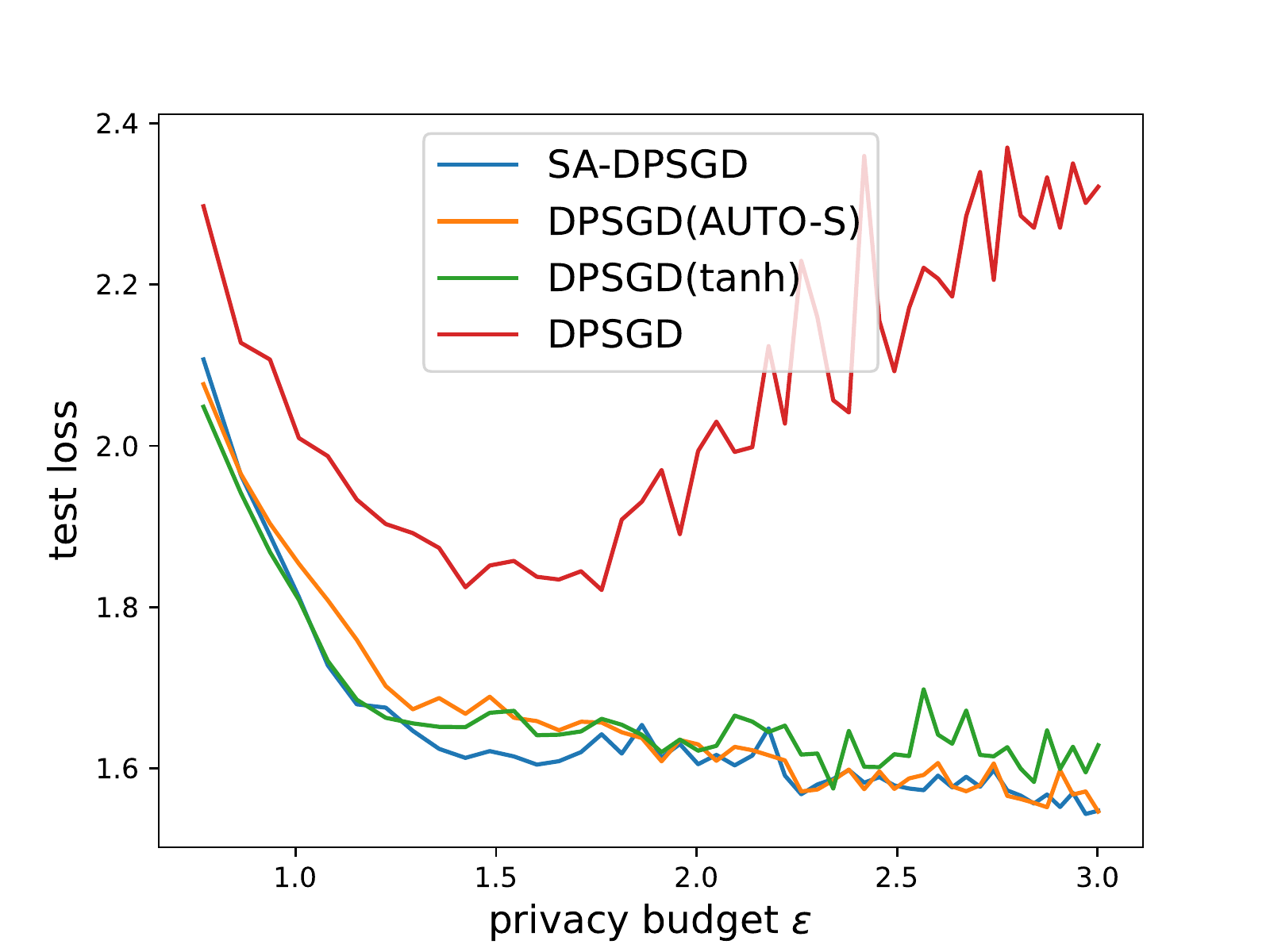}
		\caption{Avg test loss result in CIFAR10}
	\end{subfigure}
    \centering
	\begin{subfigure}{0.32\linewidth}
		\centering
		\includegraphics[width=1.0\linewidth]{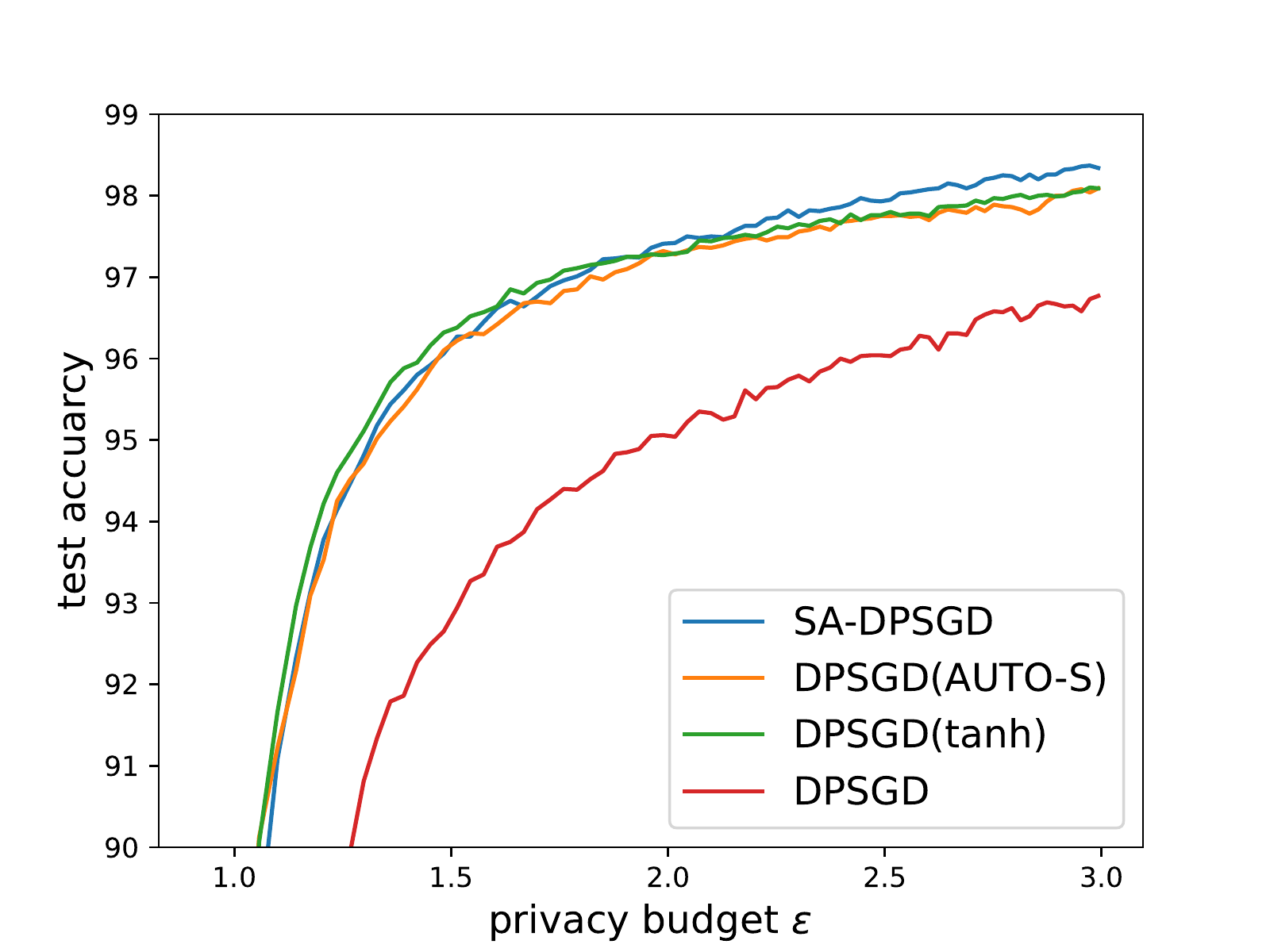}
		\caption{Avg test acc result in MNIST}
	\end{subfigure}
	\centering
	\begin{subfigure}{0.32\linewidth}
		\centering
		\includegraphics[width=1.0\linewidth]{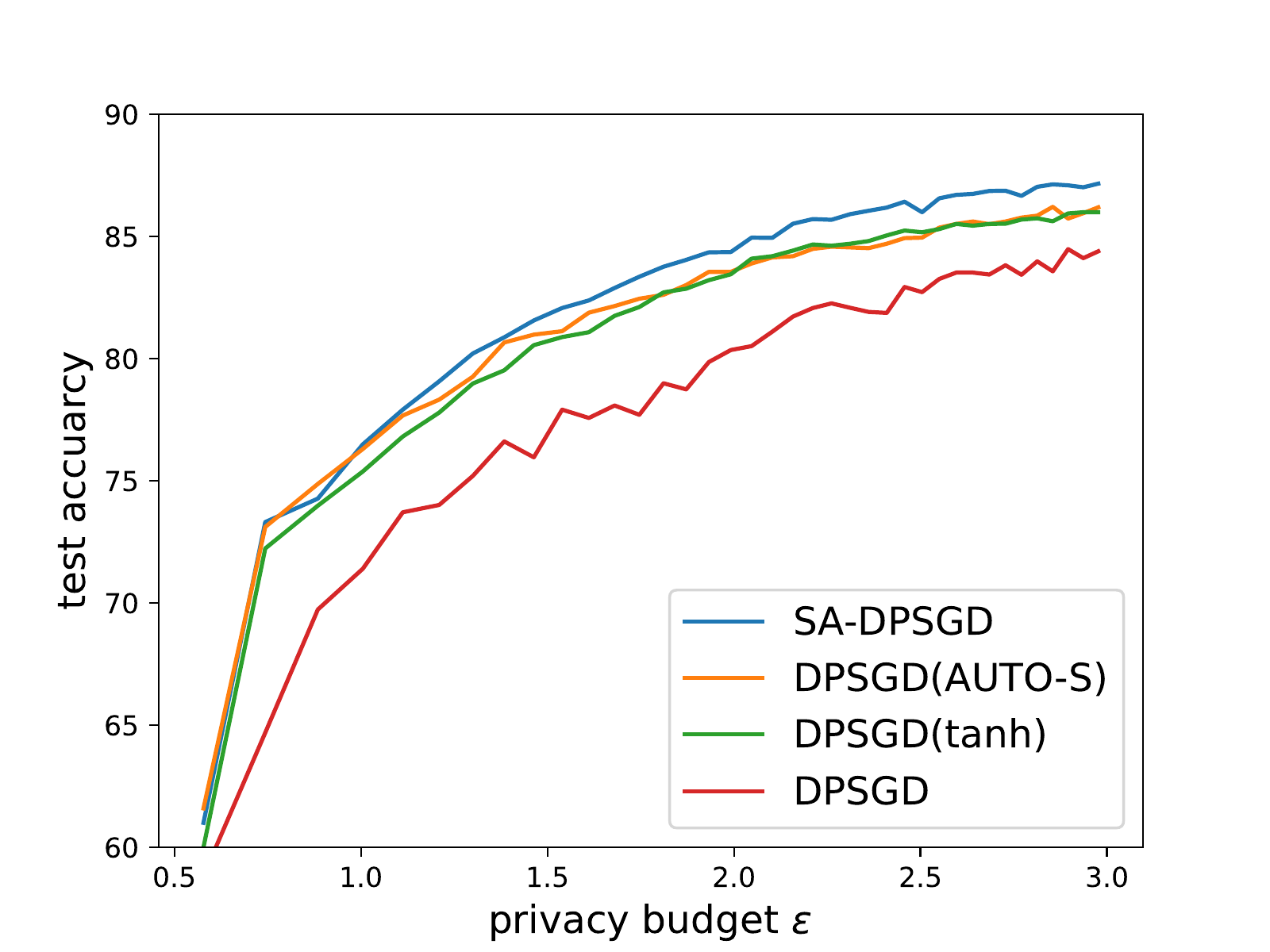}
		\caption{Avg test acc result in FashionMMNSIT}
	\end{subfigure}
	\centering
	\begin{subfigure}{0.32\linewidth}
		\centering
		\includegraphics[width=1.0\linewidth]{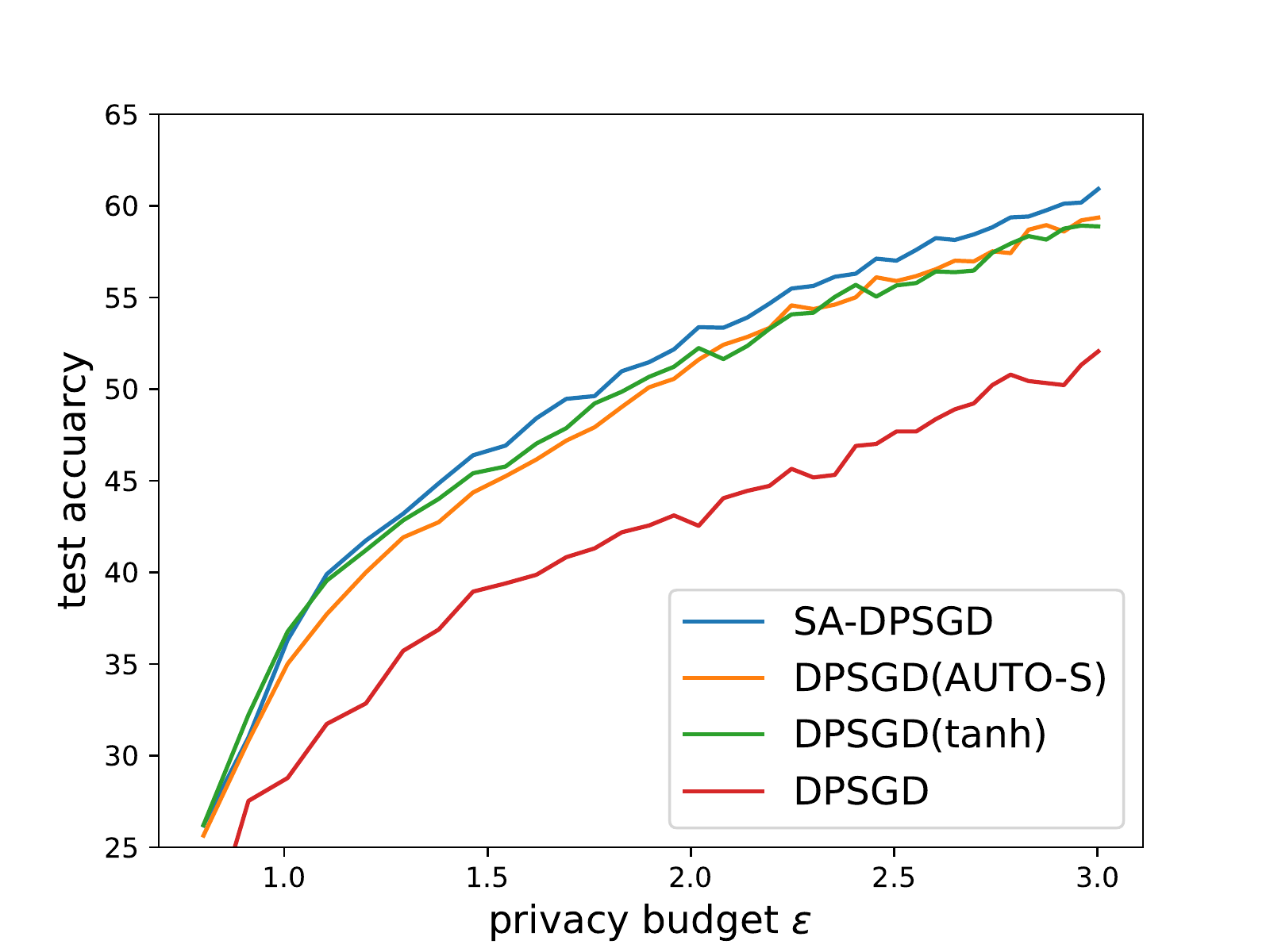}
		\caption{Avg test acc result in CIFAR10}
		\end{subfigure}
	\caption{Test average loss and accuracy as a function of the privacy loss when training with SA-DPSGD, DPSGD(AUTO-S), DPSGD(tanh) and DPSGD on MNIST, FashionMNIST,
and CIFAR10 (left to right). All elements of the architecture (number, type, and dimension of layers) and the hyperparameters (learning rate, batch size, clipping norm and noise scale) are identical. Results averaged over 5 runs.}
\label{fig:3}
\end{figure*}


We compare the utility of the four schemes when achieving the same privacy level. Fig.~\ref{fig:3} shows the privacy-utility Pareto curves\cite{BrendanAvent2019AutomaticDO} for the DPSGD, DPSGD(tanh), DPSGD(AUTO-S) and SA-DPSGD trained on all three datasets. Fig.~\ref{fig:3}(a-c) shows the test loss and privacy loss curves, we can see that SA-DPSGD almost always outperforms the previous schemes in terms of loss values. Fig.~\ref{fig:3}(d-f) shows the accuracy and privacy loss curves, and again, SA-DPSGD almost always outperforms other schemes. In particular, we can see that the larger the privacy budget is, the better SA-DPSGD outperforms the previous schemes. The reason is that our scheme selects model updates that are beneficial to model convergence by simulated annealing. This allows the model to further approach convergence in the late iterations and also saves the privacy budget. The previous schemes cannot avoid the effect from random noise, so they are difficult to further converge in the late iterations.


\begin{figure*}[htbp]
	\centering
	\begin{subfigure}{0.24\linewidth}
		\centering
		\includegraphics[width=1.0\linewidth]{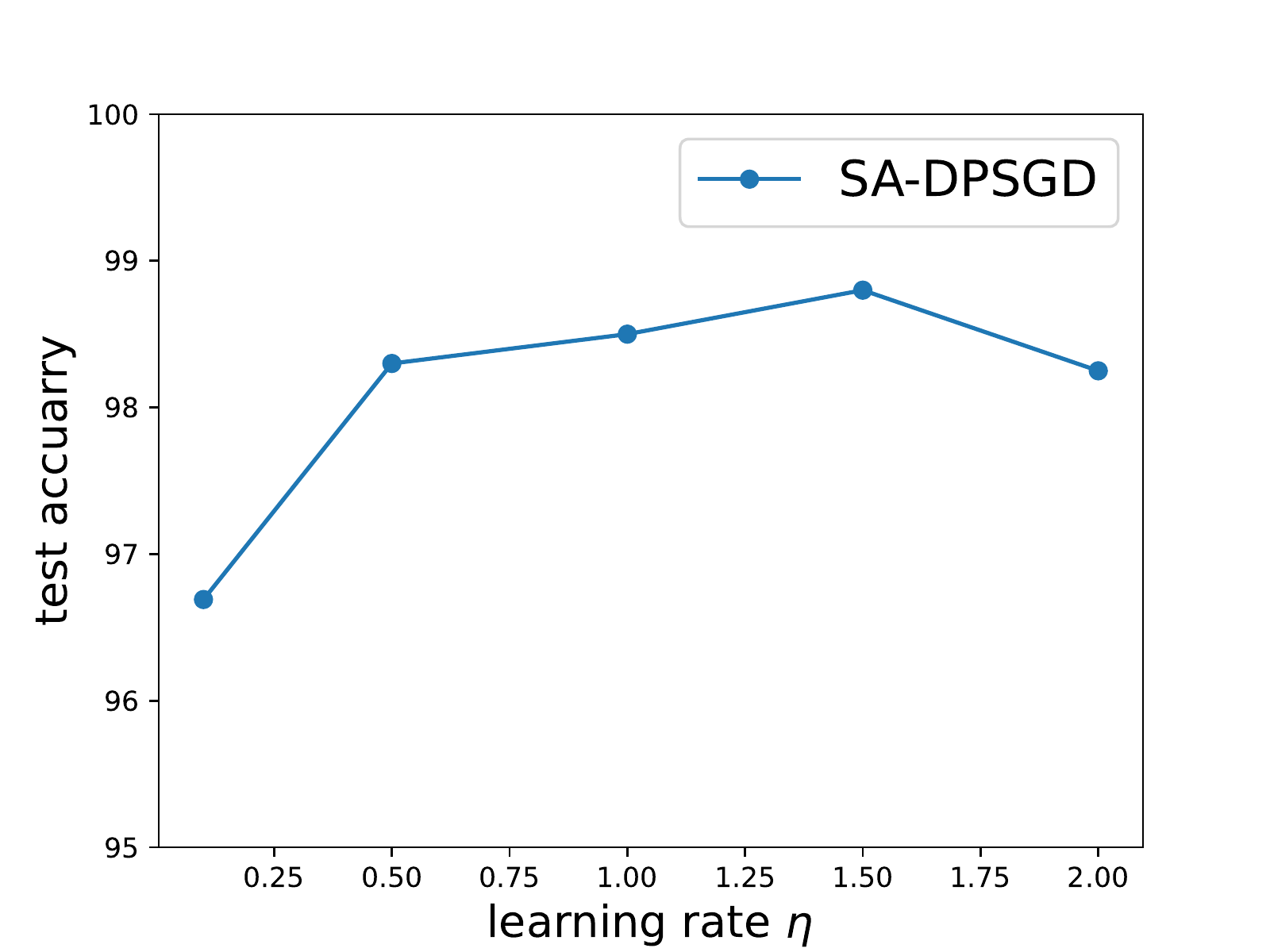}
        \label{figure:4a}
		\caption{Impact of learning rate}
	\end{subfigure}
	\centering
	\begin{subfigure}{0.24\linewidth}
		\centering
		\includegraphics[width=1.0\linewidth]{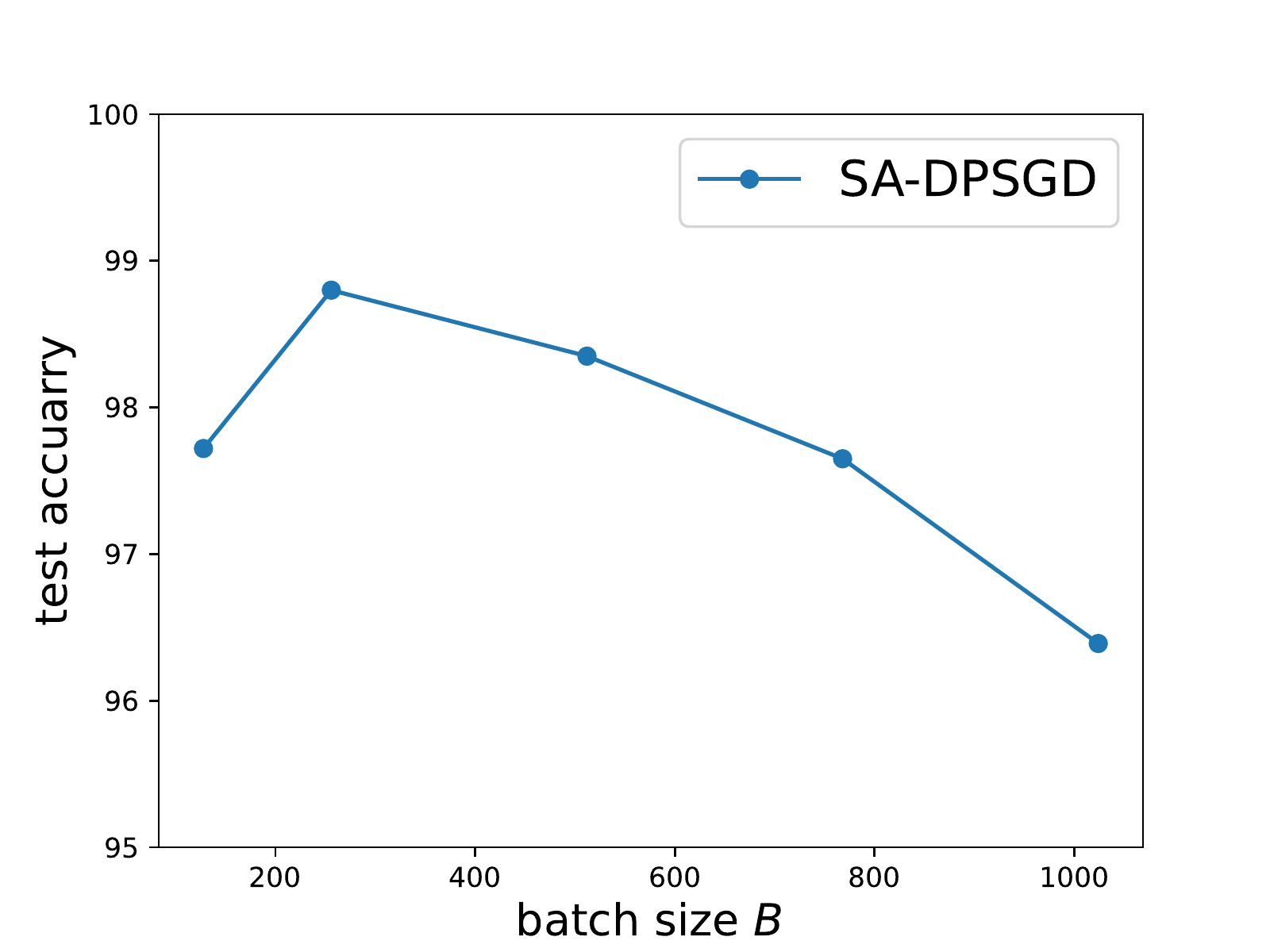}
        \label{figure:4b}
		\caption{Impact of batch size}
	\end{subfigure}
	\centering
	\begin{subfigure}{0.24\linewidth}
		\centering
		\includegraphics[width=1.0\linewidth]{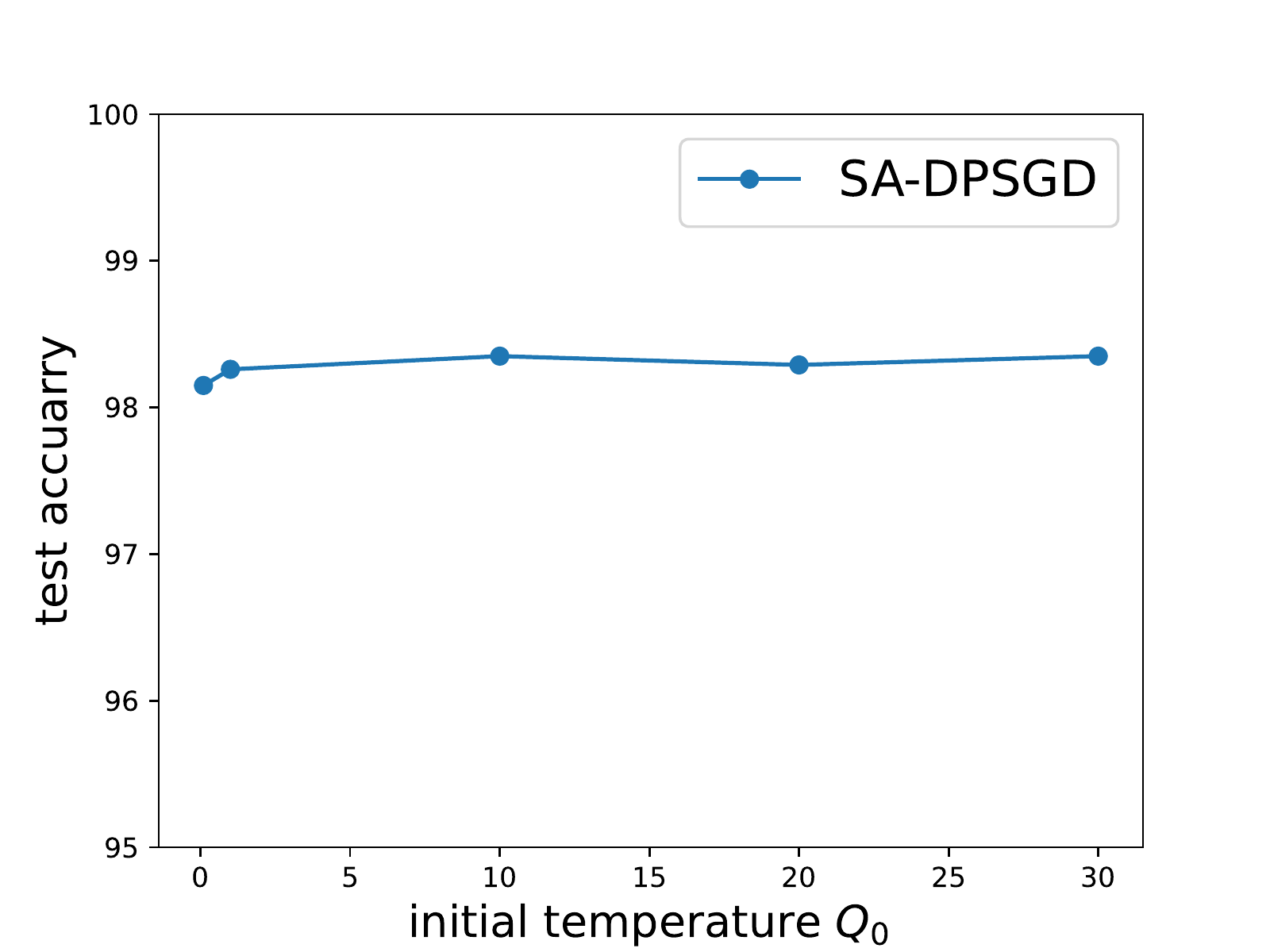}
        \label{figure:4c}
		\caption{Impact of initial temperature}
	\end{subfigure}
    \centering
	\begin{subfigure}{0.24\linewidth}
		\centering
		\includegraphics[width=1.0\linewidth]{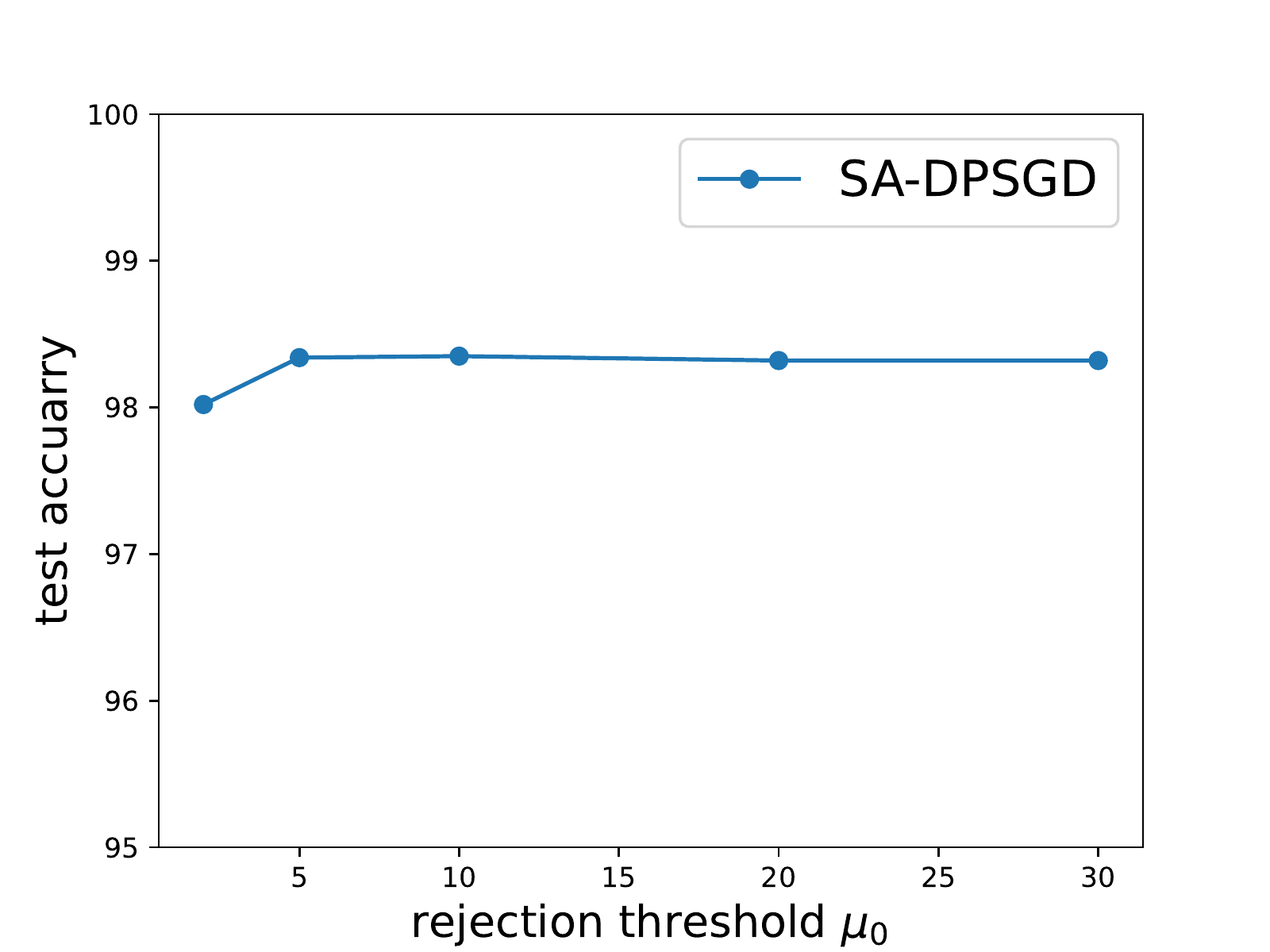}
        \label{figure:4d}
		\caption{Impact of rejection threshold}
	\end{subfigure}
	\caption{The impact of different parameters on the test accuracy of SA-DPSGD.}
\label{figure:4}
\end{figure*}

Table~\ref{tab:comp} summarizes the results with hyperparameters provided in previous work for each of the three datasets considered in our experiments.
In their own best settings and network model, the SA-DPSGD continues to consistently outperform the state-of-the-art schemes \cite{MartnAbadi2016DeepLW,NicolasPapernot2020TemperedSA,ZhiqiBu2022AutomaticCD}. For a privacy guarantee of $(\epsilon,\delta)=(3,10^{-5})$, compared to DPSGD(AUTO-S)\cite{ZhiqiBu2022AutomaticCD} which is currently the best scheme to our knowledge, our scheme achieves a test accuracy of 98.35\% on MNIST (instead of 98.12\%), 87.41\% on FashionMNIST (instead of 86.33\%), and 60.92\% on CIFAR10 (instead of 59.34\%). Furthermore, if we select hyperparameters freely, our scheme can reach even higher accuracies 98.89\%, 88.50\% and 64.17\% in the above three datasets.

\subsection{Impact of Parameters}
In this set of experiments, we study the impact of the learning rate, the batch size, the initial temperature and the rejection threshold on the MNIST dataset. In all experiments, if not specified, we use the SGD optimizer, and set the learning rate $\eta= 0.5$, batch size $B = 512$, initial temperature $Q_0 = 10$, rejection threshold $\mu_0 = 10$, noise scale $\sigma= 1.23$, and privacy budget of $(3,1e-5)$.

\textbf{Learning Rate.}
As shown in Fig.~\ref{figure:4}a, the test accuracy stays consistently above 96.5\%, irrespective of the learning rate ranging from 0.1 to 2.0. When the learning rate is lower than 1.5, the accuracy increases with the learning rate. When it is higher than 1.5, a higher learning rate results in a lower accuracy. In particular, when our method is set to a learning rate of 1.5, the test accuracy can reach 98.81\%.

\textbf{Batch Size.}
The batch size impacts on the sampling ratio, a large batch size yields a higher sampling ratio and reduces the number of training steps. A small batch size will lead to more Gaussian noise effects so that making the model fail to converge. Fig.~\ref{figure:4}b depicts the accuracy versus batch size. It can be observed that as the batch size grows, the performance (test accuracy reach 98.89\%) peaks at batch size equal to 256, and finally declines.

\textbf{Initial Temperature.}
The initial temperature $Q_0$ determines the probability of accepting an inferior solution each time, and the larger the initial temperature is, the less likely it is to accept an inferior solution. But too high an initial temperature will often put the model into a saddle point. From the Fig.~\ref{figure:4}c, we can see that when $Q_0>1$, the accuracy does not decrease and the difference is not significant.

\textbf{Rejection Threshold.}
The rejection threshold $\mu_0$ indicates the number of times our algorithm can continuously reject inferior solutions. A higher rejection threshold means more consecutive rejections of inferior solutions, and vice versa.
Fig.~\ref{figure:4}d shows that the relation of performance and rejection threshold $\mu_0$. After $\mu_0$ reaches 5, the accuracy rate remains basically the same. So a high $\mu_0$ does not help us to get higher accuracy, but on the contrary, it increases the training time.

\section{CONCLUSION}



We have proposed a differentially private scheme for stochastic gradient decent based on simulated annealing, SA-DPSGD. Our scheme uses the idea of simulated annealing to select right model updates that facilitate the convergence  during the iterations. This saves the cost of privacy budget, and finally results in more accurate models. Extensive experiments have shown that our scheme outperforms significantly the state-of-the-art schemes, DPSGD, DPSGD(tanh) and DPSGD(AUTO-S) in term of privacy cost or test accuracy. At the same time, our scheme is widely applicable and can be applied to any dataset and any neural network structure. It can serve as a strong canonical baseline for evaluating proposed improvements over DPSGD in the future. However, our scheme is time-consuming, and the future work is to investigate the choice of newly introduced hyperparameters in SA-DPSGD to trade off test accuracy and computation time.


\begin{thebibliography}{10}\itemsep=-1pt

\bibitem{MartnAbadi2016DeepLW}
Mart{\'i}n Abadi, Andy Chu, Ian Goodfellow, H.~Brendan McMahan, Ilya Mironov,
  Kunal Talwar, and Li Zhang.
\newblock Deep learning with differential privacy.
\newblock {\em Computer and Communications Security}, 2016.

\bibitem{AndresJAnayaIsaza2021AnOO}
Andres~J. Anaya-Isaza, Leonel Mera-Jim{\'e}nez, and Martha Zequera-Diaz.
\newblock An overview of deep learning in medical imaging.
\newblock {\em Informatics in Medicine Unlocked}, 2021.

\bibitem{BrendanAvent2019AutomaticDO}
Brendan Avent, Javier Gonz{\'a}lez, Tom Diethe, Andrei Paleyes, and Borja
  Balle.
\newblock Automatic discovery of privacy-utility pareto fronts.
\newblock {\em arXiv: Machine Learning}, 2019.

\bibitem{EugeneBagdasaryan2019DifferentialPH}
Eugene Bagdasaryan, Omid Poursaeed, and Vitaly Shmatikov.
\newblock Differential privacy has disparate impact on model accuracy.
\newblock {\em Neural Information Processing Systems}, 2019.

\bibitem{balle2020hypothesis}
Borja Balle, Gilles Barthe, Marco Gaboardi, Justin Hsu, and Tetsuya Sato.
\newblock Hypothesis testing interpretations and renyi differential privacy.
\newblock In {\em International Conference on Artificial Intelligence and
  Statistics}, pages 2496--2506. PMLR, 2020.

\bibitem{ZhiqiBu2022AutomaticCD}
Zhiqi Bu, Yu-Xiang Wang, Sheng Zha, and George Karypis.
\newblock Automatic clipping: Differentially private deep learning made easier
  and stronger.
\newblock 2022.

\bibitem{NicholasCarlini2019TheSS}
Nicholas Carlini, Chang Liu, {\'U}lfar Erlingsson, Jernej Kos, and Dawn Song.
\newblock The secret sharer: Evaluating and testing unintended memorization in
  neural networks.
\newblock {\em Usenix Security Symposium}, 2019.

\bibitem{AndaCheng2021DPNASNA}
Anda Cheng, Jiaxing Wang, Xi~Sheryl Zhang, Qiang Chen, Peisong Wang, and Jian
  Cheng.
\newblock Dpnas: Neural architecture search for deep learning with differential
  privacy.
\newblock {\em national conference on artificial intelligence}, 2021.

\bibitem{cummings2018role}
Rachel Cummings and Deven Desai.
\newblock The role of differential privacy in gdpr compliance.
\newblock In {\em FAT’18: Proceedings of the Conference on Fairness,
  Accountability, and Transparency}, 2018.

\bibitem{LiDeng2012TheMD}
Li Deng.
\newblock The mnist database of handwritten digit images for machine learning
  research.
\newblock {\em IEEE Signal Processing Magazine}, 2012.

\bibitem{CynthiaDwork2011AFF}
Cynthia Dwork.
\newblock A firm foundation for private data analysis.
\newblock {\em Communications of The ACM}, 2011.

\bibitem{CynthiaDwork2006CalibratingNT}
Cynthia Dwork, Frank McSherry, Kobbi Nissim, and Adam Smith.
\newblock Calibrating noise to sensitivity in private data analysis.
\newblock {\em Lecture Notes in Computer Science}, 2006.

\bibitem{dwork2014algorithmic}
Cynthia Dwork, Aaron Roth, et~al.
\newblock The algorithmic foundations of differential privacy.
\newblock {\em Foundations and Trends{\textregistered} in Theoretical Computer
  Science}, 9(3--4):211--407, 2014.

\bibitem{VitalyFeldman2019DoesLR}
Vitaly Feldman.
\newblock Does learning require memorization? a short tale about a long tail.
\newblock {\em Symposium on The Theory of Computing}, 2019.

\bibitem{MattFredrikson2015ModelIA}
Matt Fredrikson, Somesh Jha, and Thomas Ristenpart.
\newblock Model inversion attacks that exploit confidence information and basic
  countermeasures.
\newblock {\em Computer and Communications Security}, 2015.

\bibitem{AdityaGolatkar2022MixedDP}
Aditya Golatkar, Alessandro Achille, Yu-Xiang Wang, Aaron Roth, Michael Kearns,
  and Stefano Soatto.
\newblock Mixed differential privacy in computer vision.
\newblock 2022.

\bibitem{tensorflowPrivacy}
Google.
\newblock Tensorflow privacy.
\newblock {\em https://github.com/tensorflow/privacy}, 2018.

\bibitem{KaimingHe2015DeepRL}
Kaiming He, Xiangyu Zhang, Shaoqing Ren, and Jian Sun.
\newblock Deep residual learning for image recognition.
\newblock {\em arXiv: Computer Vision and Pattern Recognition}, 2015.

\bibitem{GaoHuang2016DenselyCC}
Gao Huang, Zhuang Liu, Laurens van~der Maaten, and Kilian~Q. Weinberger.
\newblock Densely connected convolutional networks.
\newblock {\em Computer Vision and Pattern Recognition}, 2016.

\bibitem{ScottKirkpatrick1983OptimizationBS}
Scott Kirkpatrick, C.~D. Gelatt, and Mario~P. Vecchi.
\newblock Optimization by simulated annealing.
\newblock {\em Science}, 1983.

\bibitem{AnttiKoskela2018LearningRA}
Antti Koskela and Antti Honkela.
\newblock Learning rate adaptation for differentially private stochastic
  gradient descent.
\newblock 2018.

\bibitem{AlexKrizhevsky2009LearningML}
Alex Krizhevsky.
\newblock Learning multiple layers of features from tiny images.
\newblock 2009.

\bibitem{JaewooLee2018ConcentratedDP}
Jaewoo Lee and Daniel Kifer.
\newblock Concentrated differentially private gradient descent with adaptive
  per-iteration privacy budget.
\newblock {\em Knowledge Discovery and Data Mining}, 2018.

\bibitem{AlexanderLundervold2018AnOO}
Alexander Lundervold and Arvid Lundervold.
\newblock An overview of deep learning in medical imaging focusing on mri.
\newblock {\em Zeitschrift Fur Medizinische Physik}, 2018.

\bibitem{LucaMelis2022ExploitingUF}
Luca Melis, Congzheng Song, Emiliano~De Cristofaro, and Vitaly Shmatikov.
\newblock Exploiting unintended feature leakage in collaborative learning.
\newblock {\em IEEE Symposium on Security and Privacy}, 2022.

\bibitem{NMetropolis1953EquationOS}
N. Metropolis, Arianna~W. Rosenbluth, Marshall~N. Rosenbluth, Augusta~H.
  Teller, and Edward Teller.
\newblock Equation of state calculations by fast computing machines.
\newblock {\em Journal of Chemical Physics}, 1953.

\bibitem{IlyaMironov2017RnyiDP}
Ilya Mironov.
\newblock R{\'e}nyi differential privacy.
\newblock {\em IEEE Computer Security Foundations Symposium}, 2017.

\bibitem{IlyaMironov2019RnyiDP}
Ilya Mironov, Kunal Talwar, and Li Zhang.
\newblock R{\'e}nyi differential privacy of the sampled gaussian mechanism.
\newblock {\em arXiv: Learning}, 2019.

\bibitem{MiladNasr2018ComprehensivePA}
Milad Nasr, Reza Shokri, and Amir Houmansadr.
\newblock Comprehensive privacy analysis of deep learning: Passive and active
  white-box inference attacks against centralized and federated learning.
\newblock {\em arXiv: Machine Learning}, 2018.

\bibitem{papernot2019machine}
Nicolas Papernot.
\newblock Machine learning at scale with differential privacy in
  $\{$TensorFlow$\}$.
\newblock In {\em 2019 $\{$USENIX$\}$ Conference on Privacy Engineering
  Practice and Respect ($\{$PEPR$\}$ 19)}, 2019.

\bibitem{NicolasPapernot2020TemperedSA}
Nicolas Papernot, Abhradeep Thakurta, Shuang Song, Steve Chien, and {\'U}lfar
  Erlingsson.
\newblock Tempered sigmoid activations for deep learning with differential
  privacy.
\newblock {\em National Conference on Artificial Intelligence}, 2020.

\bibitem{LeTrieuPhong2017PrivacyPreservingDL}
Le~Trieu Phong, Yoshinori Aono, Takuya Hayashi, Lihua Wang, and Shiho Moriai.
\newblock Privacy-preserving deep learning: Revisited and enhanced.
\newblock {\em International Conference on Applications and Techniques in
  Information Security}, 2017.

\bibitem{pytorch2018pytorch}
Automatic Differentiation~In Pytorch.
\newblock Pytorch, 2018.

\bibitem{RezaShokri2015PrivacypreservingDL}
Reza Shokri and Vitaly Shmatikov.
\newblock Privacy-preserving deep learning.
\newblock {\em Allerton Conference on Communication, Control, and Computing},
  2015.

\bibitem{song2017machine}
Congzheng Song, Thomas Ristenpart, and Vitaly Shmatikov.
\newblock Machine learning models that remember too much.
\newblock pages 587--601, 2017.

\bibitem{FlorianTramr2020DifferentiallyPL}
Florian Tram{\`e}r and Dan Boneh.
\newblock Differentially private learning needs better features (or much more
  data).
\newblock {\em Learning}, 2020.

\bibitem{ZhiboWang2018BeyondIC}
Zhibo Wang, Mengkai Song, Zhifei Zhang, Yang Song, Qian Wang, and Hairong Qi.
\newblock Beyond inferring class representatives: User-level privacy leakage
  from federated learning.
\newblock {\em International Conference on Computer Communications}, 2018.

\bibitem{LiyaoXiang2019DifferentiallyPrivateDL}
Liyao Xiang, Jingbo Yang, and Baochun Li.
\newblock Differentially-private deep learning from an optimization
  perspective.
\newblock {\em International Conference on Computer Communications}, 2019.

\bibitem{HanXiao2017FashionMNISTAN}
Han Xiao, Kashif Rasul, and Roland Vollgraf.
\newblock Fashion-mnist: a novel image dataset for benchmarking machine
  learning algorithms.
\newblock {\em arXiv: Learning}, 2017.

\bibitem{ZhiyingXu2019AnAA}
Zhiying Xu, Shuyu Shi, Alex~X. Liu, Jun Zhao, and Lin Chen.
\newblock An adaptive and fast convergent approach to differentially private
  deep learning.
\newblock {\em International Conference on Computer Communications}, 2019.

\bibitem{AshkanYousefpour2021OpacusUD}
Ashkan Yousefpour, Igor Shilov, Alexandre Sablayrolles, Davide Testuggine,
  Karthik Prasad, Mani Malek, John Nguyen, Sayan Ghosh, Akash Bharadwaj,
  Jessica Zhao, Graham Cormode, and Ilya Mironov.
\newblock Opacus: User-friendly differential privacy library in pytorch.
\newblock {\em arXiv: Learning}, 2021.

\bibitem{DaYu2021DoNL}
Da Yu, Huishuai Zhang, Wei Chen, and Tie-Yan Liu.
\newblock Do not let privacy overbill utility: Gradient embedding perturbation
  for private learning.
\newblock {\em Learning}, 2021.

\bibitem{SKevinZhou2020ARO}
S.~Kevin Zhou, Hayit Greenspan, Christos Davatzikos, James~S. Duncan, Bram van
  Ginneken, Anant Madabhushi, Jerry~L. Prince, Daniel Rueckert, and Ronald~M.
  Summers.
\newblock A review of deep learning in medical imaging: Imaging traits,
  technology trends, case studies with progress highlights, and future
  promises.
\newblock {\em arXiv: Computer Vision and Pattern Recognition}, 2020.

\bibitem{YingxueZhou2020BypassingTA}
Yingxue Zhou, Zhiwei~Steven Wu, and Arindam Banerjee.
\newblock Bypassing the ambient dimension: Private sgd with gradient subspace
  identification.
\newblock {\em Learning}, 2020.

\bibitem{zhu2019deep}
Ligeng Zhu, Zhijian Liu, and Song Han.
\newblock Deep leakage from gradients.
\newblock {\em Advances in Neural Information Processing Systems}, 32, 2019.

\end{thebibliography}


\end{document}